\documentstyle[prb,twocolumn,aps]{revtex}

\input{epsf}

\renewcommand{\H}{{\cal H}}

\newcommand{\ave}[1]{\langle #1\rangle}

\renewcommand{\b}{b}
\renewcommand{\j}{j}

\begin{document}
\draft

\title{Strong Tunneling in Double-Island Structures}

\author{Teemu Pohjola$^{1,2}$, J\"urgen K\"onig$^2$, Herbert Schoeller$^2$, 
  and Gerd Sch\"on$^{1,2}$}

\address{$^1$Materials Physics Laboratory, 
Helsinki University of Technology, 02015 HUT, Finland\\
$^2$Institut f\"ur Theoretische Festk\"orperphysik, Universit\"at Karlsruhe,
76128 Karlsruhe, Germany}

\date{\today}

\maketitle

\begin{abstract}

We study the electron transport through a system of two low-capacitance 
metal islands connected in series between two electrodes. 
The work is motivated in part by  experiments on semiconducting 
double-dots, which show intriguing effects arising from coherent tunneling of 
electrons and mixing of the single-electron states across tunneling 
barriers. In this article, we show how coherent tunneling affects 
metallic systems and leads to a mixing of the {\it macroscopic} charge
states across the barriers. We apply a recently
 formulated RG approach to examine the linear 
response of the system with high tunnel conductances (up to $8e^2/h$). 
In addition we calculate the  (second order)
cotunneling contributions to the non-linear conductance. 
Our main results are that the peaks in the linear and nonlinear
conductance as a function of the gate voltage are reduced and 
broadened in an asymmetric way, as well as shifted in their positions. 
In the limit where the two islands are coupled weakly to the electrodes,
we compare to  theoretical results obtained by Golden and Halperin 
and Matveev~{\it et al.}.
In the opposite case when the two islands are coupled  more strongly
to the leads than to each other, the peaks are found to shift, 
in qualitative agreement with the recent prediction of Andrei 
{\it et al.} for a similar double-dot system which exhibits a phase 
transition.

\end{abstract}


\section{Introduction}

Electron transport through small metal islands 
displays single-electron effects such as
Coulomb-blockade and gate-voltage dependent oscillations 
of the conductance
\cite{Ave-Lik,Gra-Dev,Sch-Uebersicht,Schoeller-MET}. 
Most of these effects can be explained in terms of lowest-order 
perturbation theory~\cite{Ave-Lik}, 
but recently it has been found -- in experiment and theory -- that 
in several regimes higher-order tunneling processes need
to be taken into account.
These are the Coulomb-blockade regime, where 
sequential tunneling is exponentially supressed, and the regime close to 
resonances. 
In the former the dominant contribution to the current
is due to cotunneling processes, 
where electrons tunnel through the system via virtual intermediate states
\cite{Ave-Naz,KSS-cot}.
Close to the resonances, cotunneling and higher-order processes 
can significantly affect the linear and nonlinear conductance even though 
the lowest-order processes are {\it not} supressed 
\cite{KSS-cot,SS,KSS1,Joyezetal}.

\begin{figure}
\epsfysize=5.5cm
\centerline{\epsffile{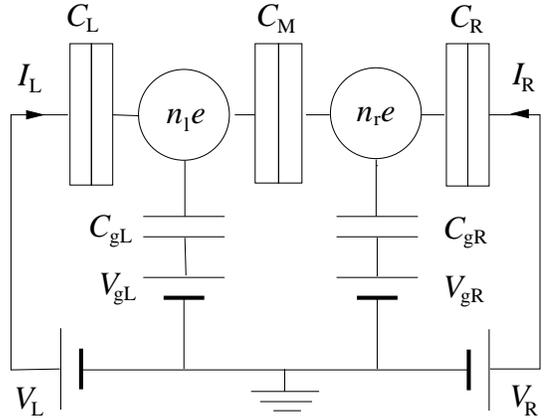}}
\caption{
  A schematic picture of a double-island structure with the voltage sources
  and various capacities in the system. The two circles denote the islands
  with $n_l/n_r$ excess electrons.
       }
\label{fg:det}
\end{figure}

In this work we study equilibrium and non-equilibrium electron transport
through the double-island system shown in Fig.~\ref{fg:det}.
Structures consisting of two coupled metal islands or large 
quantum dots (large enough to blur the discreteness of energy levels)
have recently received considerable attention
\cite{Westerveltetal,doubleislands,Mat-Gla-Bar_shift,Mat-Gla-Bar_all_alphas,Gol-Hal,Gol-Hal_largeN,AZS}.
The present work differs from the previous ones in two major respects. 
Firstly, in Refs.\ \onlinecite{Westerveltetal,doubleislands,Mat-Gla-Bar_shift,Mat-Gla-Bar_all_alphas,Gol-Hal}
and \onlinecite{AZS} dots  have been investigated which are
coupled by just one or a few conducting channels,
while in the present work a large number $N$ of parallel
channels in each junction is considered~\cite{nofchannels}.
(Ref.\ \onlinecite{Gol-Hal_largeN} also
discusses how an arbitrary number of channels in the middle junction 
affects the positions of the conductance peaks.)
Secondly, previous work has concentrated on two limiting cases: either 
strong tunneling between the two islands or between the islands and the leads, 
while we allow for arbitrary relations between the various junction 
conductances.

\begin{figure}
\epsfxsize=7.5cm
\centerline{\epsffile{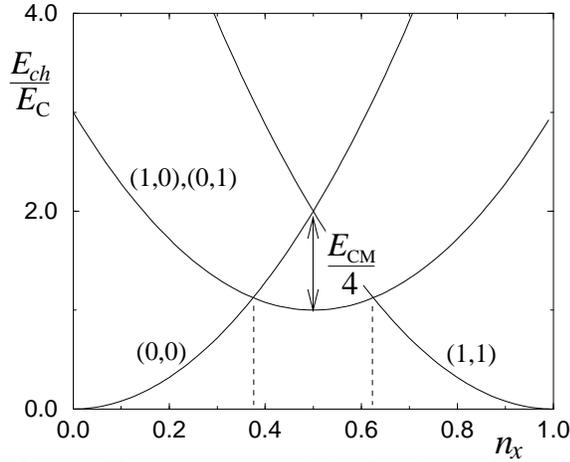}}
\caption{
  Charging energy of the four lowest-lying states for the case of equal
  gate charges $n_x=n_{x\rm l}=n_{x\rm r}$. The states are denoted 
  by (0,0), (1,0), (0,1), and (1,1).
  At temperatures $T\ll E_{\rm CM}/4$, electron transport takes place 
  in the regions around the degeneracy points of energy, depicted in 
  the figure by the vertical dashed lines.
   }
\label{fg:parabolas}
\end{figure}

\begin{figure}
\epsfxsize=6.0cm
\centerline{\epsffile{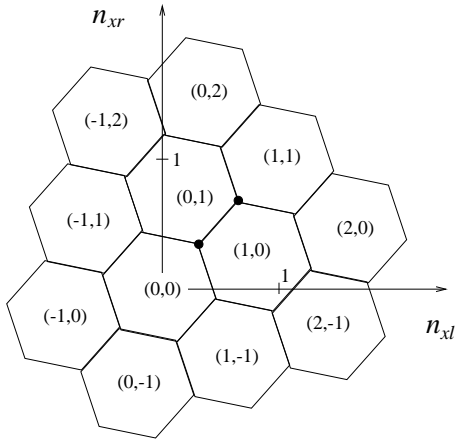}}
\caption{
  The honeycomb in the $(n_{x,\rm l},n_{x,\rm r})$-plane shows 
  the regions where a given state $(n_{\rm l},n_{\rm r})$ has the lowest energy.
  The circles denote the degeneracy points of Fig.~\ref{fg:parabolas}.
  }
\label{fg:honeycomb}
\end{figure}

Systems of tunnel junctions are described by two characteristic quantities. 
The first is the charging energy for a given
charge configuration $(n_{\rm l},n_{\rm r})$ on the islands,
\begin{eqnarray}
  E_{\rm ch}(n_{\rm l},n_{\rm r}) 
  = E_{\rm CL} (n_{\rm l} - & n_{x,{\rm l}} &)^2
          +E_{\rm CR} (n_{\rm r} - n_{x,{\rm r}})^2 \nonumber \\
     & + & E_{\rm CM} (n_{\rm l} - n_{x,{\rm l}})(n_{\rm r} - n_{x,{\rm r}}).
\label{eq:Ech}
\end{eqnarray}
It is displayed in Fig.~\ref{fg:parabolas} for the four lowest states. 
The coefficients $E_{{\rm C}b}$, with $b={\rm L},\;{\rm M},\;{\rm R}$, 
are the appropriate capacitive energy scales, which in general depend on all 
capacitances. The gate charges
$n_{x,{\rm l}}=V_{g\rm L}C_{g\rm L}/e+V_{\rm L}C_{\rm L}/e$ 
and $n_{x,{\rm r}}=V_{g\rm R}C_{g\rm R}/e+V_{\rm R}C_{\rm R}/e$
control the charge on the islands, see Fig.~\ref{fg:honeycomb}.
For convenience, we consider below a left-right symmetric situation:
$C_{\rm L}=C_{\rm R}=C$, $C_{\rm gL}=C_{\rm gR}=C_{\rm g}$, and
$n_{x,{\rm l}}=n_{x,{\rm r}}= n_x$. 
The charging energy contains two energy scales. One,
\begin{eqnarray}
  E_{\rm C} = {e^2\over 2}{1 \over 2(C+C_{\rm g})} \nonumber 
\end{eqnarray}
depends on the external capacitances and determines the curvature of 
the parabolas. 
The other scale,
\begin{eqnarray}
  E_{\rm CM} = {e^2 \over 2} {2 C_{\rm M}\over 
    (C + C_{\rm g})(C + 2 C_{\rm M} + C_{\rm g})}, \nonumber
\end{eqnarray} 
also depends on the capacitive coupling, $C_{\rm M}$, of the two islands
and leads to a splitting as displayed in Fig.~\ref{fg:parabolas}.
The other important set of parameters are the dimensionless tunneling 
conductances through the junctions $b$
\begin{equation}
  \alpha_0^b = {1\over 4\pi^2} {h \over e^2 R_{{\rm T},b}}.
\end{equation}
Also here we choose left-right symmetry, $\alpha_0^{\rm L}=\alpha_0^{\rm R}$,
but do not fix the ratio 
between $\alpha_0^{\rm M}$ and $\alpha_0^{\rm L,R}$.

In this article we study the system using two approaches.
Firstly, we make use of the real-time diagrammatic technique developed 
for metal islands \cite{KSS-cot,SS,KSS1} and quantum dots \cite{KSSS}.
This approach is applicable for linear and nonlinear response and all 
values of the gate voltages.
In the main text, we discuss a systematic perturbative expansion of 
the diagrams and, in appendix \ref{app:rates}, present the calculation 
done up to second-order in the tunneling conductances.
This calculation accounts for all second-order tunneling processes
and turns out to be independent of any cutoffs.
Secondly, in order to consider stronger tunneling, 
we study the double-island system in terms of 
a recently-devoloped renormalization-group approach \cite{KS-RG}. 
In its general form, this procedure is cutoff independent and allows us 
to include all charge states. It thus differs from the RG approach developed 
in Ref.\ \onlinecite{Matveev-RG}, which only uses two adjacent
charge states of the single-electron box and which requires a
cutoff of the order of the charging energy $E_{\rm C}$.
We apply the RG equations in their equilibrium form,
which is sufficient for the linear conductance close to resonances,
and consider tunneling conductances up to $8 e^2/h$
($\alpha_0^b$'s up to 0.20).

Our main results are the following.
In the weak-tunneling limit, $\alpha_0^b\ll1$, and low temperature 
$T\ll E_{\rm CM}/4$ the linear conductance $G$ displays 
a series of peaks as a function of the gate voltage. 
As the tunneling conductances $\alpha_0^b$ are increased, 
the peak heights are reduced with a temperature dependence
resembling the functional form $1/[A+B\log(T/E_{\rm C})]$ ($A$ and $B$ are 
$T$-independent parameters).
Also the peak positions are renormalized due to higher-order processes.
For the special case when the two islands are only weakly coupled to the leads,
i.e., $\alpha_0^{\rm M}\gg\alpha_0^{\rm L,R}$, we reproduce the 
$\alpha_0^{\rm M}$-dependences of the peak positions 
found in Refs.~\onlinecite{Mat-Gla-Bar_shift} and \onlinecite{Gol-Hal_largeN}.
As new features we find the corrections arising due to $\alpha_0^{\rm L,R}$ 
including a weak temperature dependence of the peak positions.
In the opposite limit, $\alpha_0^{\rm L,R}\gg\alpha_0^{\rm M}$ and 
just one channel in each junction, 
Andrei {\it et al.}\cite{AZS} predict a phase transition:
the two conductance peaks around $n_x=0.5$ 
coalesce into one as $E_{\rm CM}$ is reduced below a finite critical 
value $E_{\rm CM}^{\rm cr}$. 
By putting $\alpha_0^{\rm M}\ll\alpha_0^{\rm L,R}$ 
we observe a similar trend for our junctions with large
number of channels $N$.

For completeness, we use the diagrammatic technique to account for 
the nonlinear conductance ($dI/dV\;{\rm at}\;V\not=0$) and the inelastic 
cotunneling processes, which cannot be studied with the equilibrium 
RG procedure. 
The nonlinear conductance displays a double-peak structure resembling 
that found for SET's; in the present case the peaks are typically asymmetric.
Second-order processes enhance this asymmetry and pull the two peaks further 
apart. On top of this, the inelastic cotunneling processes result in 
algebraically decaying tails of the peaks both in linear and nonlinear 
conductance.

\pagebreak

\section{Microscopic model and sequential tunneling}

The double-island system is described by the Hamiltonian
\begin{eqnarray}
        \H   & = & \H_0 + \H_{\rm T} {\rm \;\;\;\; with} \nonumber\\
        \H_0 & = & \H_{\rm L} + \H_{\rm R} + \H_{\rm l} + \H_{\rm r} + 
        \H_{\rm ch}.
\label{eq:Hamiltonian}
\end{eqnarray}
The terms $\H_\j$ describe noninteracting electrons in the leads and islands:
$\H_\j=\sum_{k,n}\varepsilon_{kn}^\j a_{\j kn}^\dagger 
a_{\j kn}$ with $\j\in\{{\rm L,R}\}$ and
$\H_i=\sum_{q,n}\varepsilon_{qn}^i c_{iqn}^\dagger 
c_{iqn}$ with $i\in\{{\rm l,r}\}$.
The indices $k$ and $q$ label the electronic states, and $n=1\ldots N$ labels
the transverse channels.
We assume that the number of channels $N$ in each junctions is large.
The Coulomb interactions are described by the term $\H_{\rm ch}$ which
is the operator form of Eq.~(\ref{eq:Ech}).
Tunneling of electrons is described in terms of the tunneling Hamiltonian 
\begin{eqnarray}
  \H_{\rm T} & = & \sum_{kqn} \left( T_{kq}^{{\rm L}n} 
    a_{{\rm L}kn}^\dagger c_{{\rm l}qn} e^{-i\hat{\phi}_{\rm l}}
    +\!{\rm H.c.}\right) \nonumber \\
  & + & \sum_{kqn} \left( T_{kq}^{{\rm R}n} 
    a_{{\rm R}kn}^\dagger c_{{\rm r}qn} e^{-i\hat{\phi}_{\rm r}}
    +\!{\rm H.c.}\right) \nonumber \\  
  & + & \sum_{qq^\prime n} \left( T_{qq^\prime}^{{\rm M}n} 
    c_{{\rm l}qn}^\dagger c_{{\rm r}q^\prime n} 
    e^{i(\hat{\phi}_{\rm l} - \hat{\phi}_{\rm r})}
    +\!{\rm H.c.}\right),
\end{eqnarray}
where $T_{kq}^{bn}$ are the tunneling matrix elements for the barriers $b$
(below we assume $T_{kq}^{bn}=T^{bn}$).
They are related to the tunneling resistances, e.g.,
for the right junction via
\begin{equation}
  \alpha_0^{\rm R} = {1\over 4\pi^2} {h \over e^2R_{\rm T,R}} =
  \sum_n N_{\rm R}(0) N_{\rm r}(0) \left|T^{{\rm R},n}\right|^2.
\label{eq:alphas}
\end{equation}
Here $N_{\rm R/r}(0)$ is the density of states in the right lead/island
and the operator $e^{\pm i\hat{\phi}_i}$ changes the charge on the island $i$ 
by $\pm e$.

Due to the periodicity of energies, it is sufficient to concentrate on 
gate voltages in the range $0\le n_x\le 0.5$. 
In this interval, the tunneling of electrons 
through the barrier $b={\rm L,M,R}$ is characterized by the following 
differences in 
charging energies:
$\Delta_{\rm L}\equiv E_{ch}(1,0)-E_{ch}(0,0)$, 
$\Delta_{\rm R}\equiv E_{ch}(0,1)-E_{ch}(0,0)$, and 
$\Delta_{\rm M}\equiv E_{ch}(0,1)-E_{ch}(1,0)$.

In the weak tunneling limit, $\alpha_0\ll1$, sequential tunneling 
processes dominante.
At low temperatures only the states $(0,0)$, $(1,0)$, and $(0,1)$ participate in
the transport and we obtain the following expression for the current
\begin{eqnarray}
  I^{\rm seq} && =  {2\pi e \over \hbar}
\label{eq:Iseq} \\
  &&      \times         
                 { {\alpha_{\rm L}\alpha_{\rm M}\alpha_{\rm R}
                   (f_{\rm L}^+f_{\rm M}^+f_{\rm R}^-
                   - f_{\rm L}^-f_{\rm M}^-f_{\rm R}^+)}
                 \over
                 {\alpha_{\rm M}^+\alpha_{\rm R} 
                   + \alpha_{\rm M}^-\alpha_{\rm L}
                   + \alpha_{\rm M}^+\alpha_{\rm L}^+
                   + \alpha_{\rm M}^-\alpha_{\rm R}^+
                   + \alpha_{\rm R}^-\alpha_{\rm L}
                   + \alpha_{\rm R}^+\alpha_{\rm L}^-}}.\nonumber
\end{eqnarray}
For compactness of this expression, we have omitted the argument $\Delta_b$ 
from both the sequential-tunneling rates
\begin{equation}
        \alpha_\b^\pm(\Delta_b) = \pm\alpha_0^\b
        {{\Delta_b-\Delta\mu_\b} 
        \over {\exp\left[\pm\beta(\Delta_b-\Delta\mu_\b)\right] - 1}}
\label{eq:seqtunrate}
\end{equation}
for the barriers $b$ as well as the Fermi functions 
$f_b^\pm(\Delta_b)=(\exp[\pm\beta(\Delta_b-\Delta\mu_b)]+1)^{-1}$ 
($\Delta\mu_b$ is the difference in chemical potentials across barrier $b$).
We further introduced 
$\alpha_b\equiv\alpha_b^+(\Delta_b) + \alpha_b^-(\Delta_b)$.

\begin{figure}[h]
\epsfxsize=8cm
\centerline{\epsffile{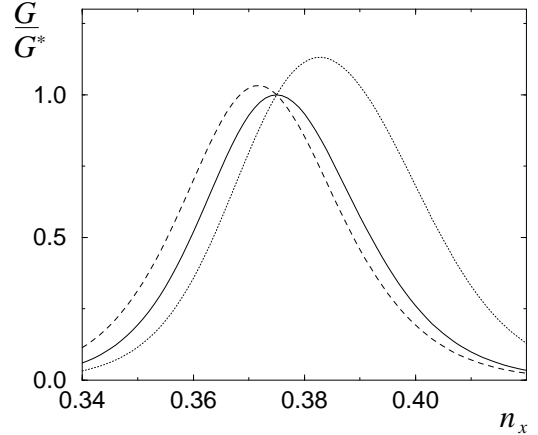}}
\caption{
  Sequential-tunneling conductance for junction conductances 
  $\alpha_0^{\rm M}= \alpha_0^{\rm L,R}=0.01$ (solid line), 
  $\alpha_0^{\rm M}=0.001,\;\alpha_0^{\rm L,R}=0.01$ (dotted line), 
  and $\alpha_0^{\rm M}=0.01,\;\alpha_0^{\rm L,R}=0.001$ (dashed line).
  The curves intersect at the degeneracy point of the charging energies 
  due to scaling with Eq.~(\ref{eq:Gseq}).  
  For $\alpha_0^{\rm M}$ larger (smaller) than $\alpha_0^{\rm L,R}$ 
  increasing temperature shifts the peak to the left (right).
  }
\label{fg:Gseq}
\end{figure}

Already the sequential-tunneling current, Eq.~(\ref{eq:Iseq}), displays 
interesting properties not found in single-island transistors.
In the linear conductance regime one would expect to find the conductance 
peak at the degeneracy point, and this is indeed the case
when all $\alpha_0^b$ are equal. In the asymmetric case, 
$\alpha_0^{\rm M}\neq\alpha_0^{\rm L,R}$, the peak acquires a 
temperature-dependent shift (vanishing at $T=0$).
The direction of the shift depends on the relative 
magnitude of $\alpha_0^M$ and $\alpha_0^L=\alpha_0^R$
as shown in Fig.~\ref{fg:Gseq}.
Remarkably, the linear conductance at the degeneracy point is always 
given by the temperature-independent expression
\begin{equation}
  G^* = {2\pi\over 3}{e^2\over\hbar}
            {\alpha_0^{\rm L} \alpha_0^{\rm M} \alpha_0^{\rm R} \over
             \alpha_0^{\rm L} \alpha_0^{\rm M} 
             + \alpha_0^{\rm L} \alpha_0^{\rm R} 
             + \alpha_0^{\rm M} \alpha_0^{\rm R}}
\label{eq:Gseq}
\end{equation}
corresponding to the intersection point, $n_x=0.375$, in Fig.~\ref{fg:Gseq}.

In what follows, we account for charge fluctuations through all three 
junctions. 
We do this using two complementary approaches: a real-time diagrammatic 
technique and a renormalization-group analysis.

\pagebreak

\section{Diagrammatic approach}

The physical properties, including the nonequilibrium dynamics of 
the double-island system, are described by a reduced density matrix.
We study the time evolution of this density matrix in a basis of 
charge states, $\chi=(n_{{\rm l},\chi},n_{{\rm r},\chi})$, and 
perform an expansion in $\H_{\rm T}$. 
This leads to a formally exact master equation for the occupation 
probabilities $p_\chi$ in terms of general transition rates 
$\Sigma_{\chi,\chi^\prime}$ which remain to be evaluated.
In the stationary limit the master equation reads \cite{SS,KSS1}
\begin{equation}
        0=\dot{p}_\chi = \sum_{\chi^\prime\neq \chi}
                \left[p_{\chi^\prime}\Sigma_{\chi^\prime,\chi}
                - p_\chi\Sigma_{\chi,\chi^\prime}\right].
\label{eq:master}
\end{equation}
Also the stationary current through the junction $b$ can be expressed 
in terms of rates and probabilities,
\begin{equation}
        \ave{I_b} = -ie\sum_{\chi,\chi^\prime}p_{\chi^\prime}
                        \Sigma_{\chi^\prime,\chi}^{b+}.
\label{eq:current}
\end{equation}
The rate $\Sigma_{\chi^\prime,\chi}^{b+}$ with the plus sign is a subset
of all possible rates contained in $\Sigma_{\chi,\chi^\prime}$ \cite{sigmaplus}.
We refer the reader to Refs.~\onlinecite{Schoeller-MET} and \onlinecite{KSSS} 
for a more thorough discussion on the above equation.

\begin{figure}[h]
\epsfxsize=8.5cm
\centerline{\epsffile{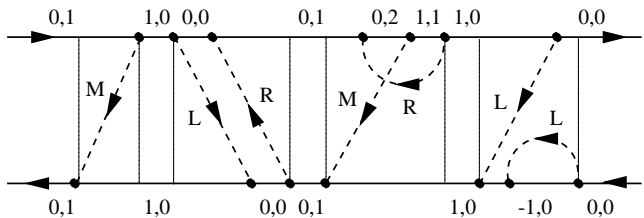}}
\caption{
  Example of the diagrammatic representation of the time evolution of 
  the reduced density matrix. In the diagrams one can distinguish 
  different tunneling processes (the vertical dotted lines are meant 
  to guide the eye), from left to right,    
  sequential tunneling, inelastic cotunneling, and second-order
  processes leading to vertex and propagator corrections. 
  The pairs of numbers on the forward and backward propagators (the horizontal
  lines) correspond to charge states, the dashed lines with labels $b$
  describe tunneling of electrons across junction $b$. 
  }
\label{fg:diagram}
\end{figure}

The nonequilibrium time evolution of the density matrix 
 may be visualized in terms of diagrams.
In particular,
the rate $\Sigma_{\chi,\chi^\prime}$ is a sum of all possible processes
of which Fig.~\ref{fg:diagram} shows some examples.
In previous work, two approaches have been used for evaluating the rates.
In the so-called resonant-tunneling approximation \cite{SS,KSS1} 
one occounts for an infinite sum of terms of a certain form
via the use of self-consistent equations in the spirit of the Dyson equation.
This approach  accounts for high-order tunneling processes but is, 
in practice, limited to a small number of charge states.
In the alternative, a systematic perturbative expansion of the diagrams 
\cite{KSS-cot}, one includes all processes of a given order $O(\alpha^k_0)$.
The key advantage of this approach is that, for not too high tunneling 
conductances, it can account for a larger number of charge states.

The probabilities and rates can be expanded in powers of $\alpha_0$ 
\cite{KSSS,KSS-cot}
\begin{eqnarray}
  p_\chi & = & p_\chi^{(0)} + p_\chi^{(1)} + p_\chi^{(2)} +\dots\nonumber\\
  \Sigma_{\chi,\chi^\prime} & = & \Sigma_{\chi,\chi^\prime}^{(1)} +
   \Sigma_{\chi,\chi^\prime}^{(2)} + \Sigma_{\chi,\chi^\prime}^{(3)} +\dots, 
\label{eq:series}
\end{eqnarray}
where $p_\chi^{(k)}$ and $\Sigma_{\chi,\chi^\prime}^{(k)}$ denote the terms
$\sim\alpha_0^k$. 
Figure \ref{fg:diagram} examplifies processes contributing to 
the rates $\Sigma^{(k)}$ with $k=1,2$.
When the expressions (\ref{eq:series}) are inserted into 
the master equation, Eq.~(\ref{eq:master}), this must hold in each order 
and once we know the rates we may iteratively solve for $p^{(k)}$.
The probabilities and rates together determine the average charge on 
the islands $\ave{n_i}=\sum_\chi n_{i,\chi}p_{i,\chi}$ as well as the current 
$I\equiv\ave{I_b}$, from Eq.~(\ref{eq:current}).

A second-order calculation $(O(\alpha_0^2))$
that accounts for all charge states (reachable in this order) 
has proven its virtues in the study of SET's: 
the calculations could be carried out exactly and all cutoff dependences 
vanish from the physical quantities. 
Also the comparison\cite{KSS-cot} with experiments 
showed that proper consideration of cotunneling processes 
gives correct results for a wide range of tunneling conductances and also 
at resonances.
A similar second-order calculation for the double-island system together 
with rules for evaluating the individual diagrams is presented in appendix 
\ref{app:rates}. Also the results for linear and nonlinear conductance 
are shown there.

\section{Renormalization group analysis}

In the strong tunneling regime, such as attained in recent experiments
on single-electron transistors\cite{Joyezetal}, it is necessary to go beyond
low-order expansions in the tunneling conductances. 
To this end, we study the double-island system using
a recently developed real-time RG procedure\cite{KS-RG}.
In its general form, this procedure accounts for an arbitrary number 
of charge states and is independent of any cutoffs. 
Below we use the equilibrium form of the RG equations which yields 
the renormalization of system parameters.
We extend the equilibrium description of the system to account for transport 
properties via the following assumption: 
for not too large $\alpha_0^b$ the linear conductance is given up
to small regular terms by the sequential-tunneling expressions when we replace 
all the bare parameters with the renormalized ones. 
The linear conductance obtained in this manner is consistent 
with the diagrammatic calculation at least up to $O(\alpha_0^2)$.
It should be noted that this approach does not account for the inelastic 
cotunneling processes. 
However, here we are mainly interested in the region around the conductance 
maximum where, at least in the order $O(\alpha_0^2)$, these only result in 
a weak temperature independent correction (similar result obtained 
for a SET).

The results from the full RG are obtained numerically and are thus presented 
graphically. For small $\alpha_0^b$ the RG flow equations can be
expanded analytically up to $O(\alpha_0^2)$ and we can complement 
the numerical results with the analytical formulas obtained in this order.
The energy differences $\Delta^{(0)}(n_x)=E_{\rm ch}(1,0;n_x)-E_{\rm ch}(0,0;n_x)=E_{\rm ch}(0,1;n_x)-E_{\rm ch}(0,0;n_x)$ 
and the tunneling conductances $\alpha_0^b$ are found to be renormalized to 
$\tilde{\Delta}(n_x)=\Delta^{(0)}(n_x)+\Delta^{(1)}(n_x)$ and
$\tilde{\alpha}^b=\alpha_0^b+\alpha_b^{(2)}$, respectively. 
The former yields directly the renormalized degeneracy point, $n_x^*$, 
following from the condition 
$\tilde{\Delta}(n_x^*)=\Delta^{(0)}(n_x^*)+\Delta^{(1)}(n_x^*)=0$,
see Ref.~\onlinecite{expansion}.
One should keep in mind that the peak arising from the expression for 
the sequential-tunneling conductance is not necessarily located at 
the degeneracy point, see Fig.~\ref{fg:Gseq}.
The peak may be shifted due to unequal $\alpha_0^b$ but also for large 
$\alpha_0^b$ because these can become significantly unequal in 
the renormalization flow.
The full RG equations as well as their $O(\alpha_0^2)$-expansion
(see below) for the general case are sketched in Appendix \ref{app:RG}.

\section{Results}
\label{sc:results}

\begin{figure}[h]
\epsfxsize=8.0cm
\centerline{\epsffile{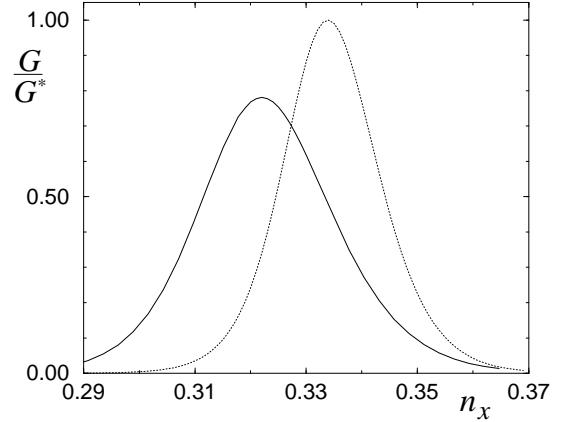}}
\caption{
  The solid line shows an example of the linear conductance peak as obtained 
  with the full RG calculation. The dotted curve is the sequential-tunneling 
  conductance for the same parameters. 
  Here $\alpha_0^{\rm L}=\alpha_0^{\rm R}=0.025$,
  $\alpha_0^{\rm M}=0.05$ and $\ln(T/E_{\rm C})=-4$.
  }
\label{fg:Gex}
\end{figure}

Figure \ref{fg:Gex} shows the main effects of increasing tunneling strength
on the linear conductance: the conductance peak is shifted and 
broadened and its maximum is reduced.
Furthermore the peak shape, in particular in the tail regions, 
becomes asymmetric. 
The key for understanding the changes around the conductance peak maximum
is the renormalization of the system parameters:
the peak lies at or close to the position of the degeneracy point, 
see Fig.~\ref{fg:parabolas}, which is determined by the capacitive energies 
$\tilde{E}_{\rm C}$ and $\tilde{E}_{\rm CM}$,
the peak width depends on the curvature of the energy parabolas,
$\tilde{E}_{\rm C}$, and the peak height on the tunneling 
conductances $\tilde{\alpha}_b$.
For increasing tunneling strength, the effective capacitances $\tilde{C}_b$ 
typically increase (modifying $E_{{\rm C}b}$ into $\tilde{E}_{{\rm C}b}$) 
while $\tilde{\alpha}_b$ decrease.
The fluctuations through the middle and the outer junctions affect the system
in qualitatively different manner and below we distinguish
between the three cases, $\alpha_0^{\rm M}\gg\alpha_0^{\rm L,R}$, 
$\alpha_0^{\rm M}\ll\alpha_0^{\rm L,R}$ and 
$\alpha_0^{\rm M}=\alpha_0^{\rm L,R}$. 
In all the figures, we have chosen left-right symmetry and, unless otherwise 
mentioned, used the convention 
$C_{\rm L}=C_{\rm R}=C_{\rm M}=100C_{\rm gL}=100C_{\rm gR}$ similar to the
experimental values of Joyez~{\it et al}\cite{Joyezetal}. 
In terms of the charging energies, this corresponds to 
$E_{\rm CM} \approx E_{\rm CL,R} \approx {4\over3} E_{\rm C}$.

\begin{figure}[h]
\epsfxsize=8.0cm
\centerline{\epsffile{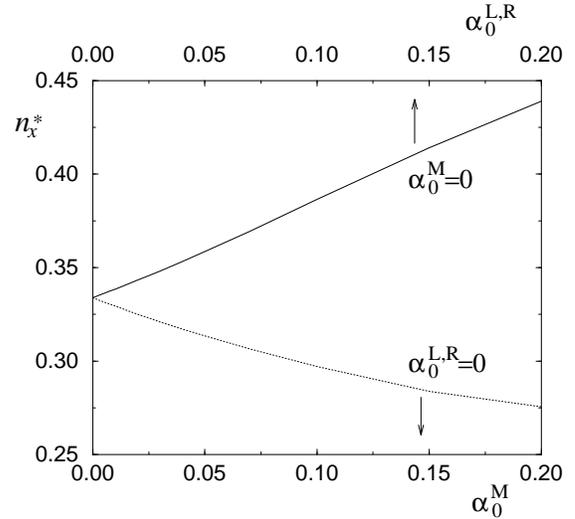}}
\caption{
  The position of the degeneracy point $n_x^*$ at $T=0$
  as a function of $\alpha_0^{\rm M}$ with $\alpha_0^{\rm L,R}=0$ 
  (dotted line, lower horizontal axis) and 
  $\alpha_0^{\rm L,R}$ with $\alpha_0^{\rm M}=0$ 
  (solid line, upper horizontal axis).
  }
\label{fg:peakshift}
\end{figure}

The case $\alpha_0^{\rm M}\gg\alpha_0^{\rm L,R}$ consists of
a subsystem, the two islands, whose properties are probed via 
the weak tunneling junctions connecting the subsystem to the leads.
For weak coupling between the islands, i.e., small $\alpha_0^{\rm M}$,
the electrons in each island form well-defined quantum states. 
For increasing $\alpha_0^{\rm M}$, the states in different islands
become increasingly mixed reducing the resulting ground-state energy. 
In terms of the system parameters,
this can be seen as an increase in $C_{\rm M}$ -- increase in $E_{\rm CM}$ --
which lowers the energy of the states (1,0) and (0,1) and shifts $n_x^*$ 
correspondingly, see Figs.~\ref{fg:peakshift} and \ref{fg:parabolas}.
The renormalization of $\alpha_0^b$'s and $\Delta$ is independent of 
the gate voltage and the peak width remains given by thermal broadening.

\begin{figure}[h]
\epsfxsize=8.0cm
\centerline{\epsffile{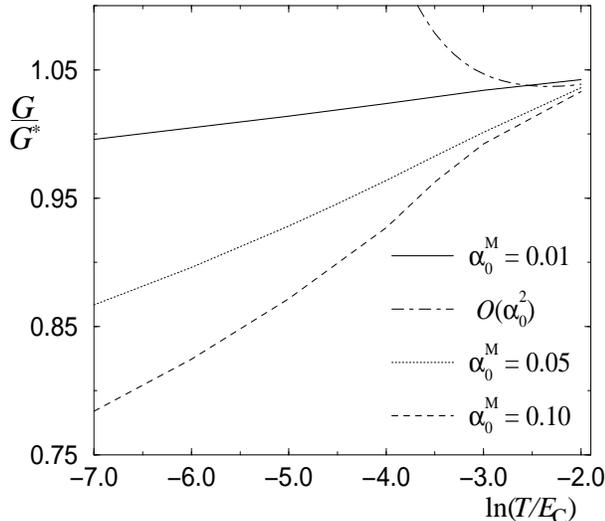}}
\caption{
  Temperature dependence of 
  the conductance maxima for $\alpha_0^{\rm M}=0.01,0.05,0.1$.
  The curves approach logarithmic form for $\alpha_0^{\rm M}\rightarrow0$.
  The upward shooting curve is the second-order result for 
  $\alpha_0^{\rm M}=0.01$, see text for further discussion on its shape.
  }
\label{fg:fullG_aM}
\end{figure}

Figure \ref{fg:fullG_aM} displays the temperature dependence 
of the maximum conductance (the curves approach logarithmic form for 
$\alpha_0^{\rm M}\rightarrow0$). It
shows also the maximum conductance for 
$\alpha_0^{\rm M}=0.01$ as obtained in second-order. 
The functional form is seen to differ from the corresponding RG result 
in three ways. First, the overall diverging shape is due to an unphysical 
$\Delta^{(1)}/T$-dependence of the perturbation expansion, 
emphasized for cases where the peak experiences
a strong shift (large $\Delta^{(1)}$). 
Second, the initial slope at higher temperatures is slightly overestimated 
by the second-order result. 
Third, the maximum conductance is shifted downwards as compared to the RG 
result. This is due to the neglect of the inelastic cotunneling contributions
in the RG calculation.

\begin{figure}[h]
\epsfxsize=7cm
\centerline{\epsffile{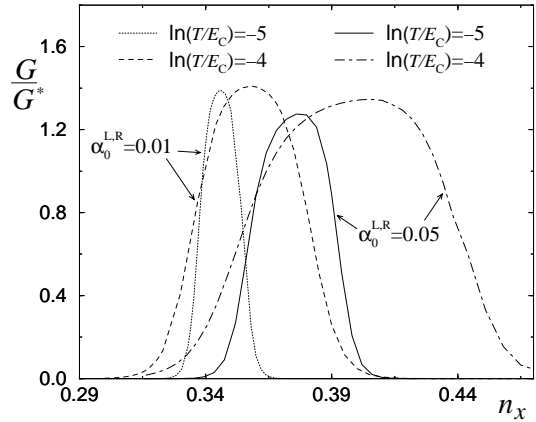}}
\caption{
  The full conductance curve at $\ln(T/E_{\rm C})=-5,-4$
  for $\alpha_0^{\rm L,R}=0.01$ (dotted and dashed, respectively) and for
  $\alpha_0^{\rm L,R}=0.05$ (solid and dot-dashed, respectively).
  The degeneracy points of energy are located at 0.3385 and 0.3586
  for $\alpha_0^{\rm L,R}=0.01,0.05$, respectively.
  }
\label{fg:fullG_aL}
\end{figure}

The limit $\alpha_0^{\rm L,R}\gg\alpha_0^{\rm M}$
corresponds to two inter-connected single-electron boxes (SEB) where 
the small conductance at the central junction can be used to probe 
the capacitive interactions of the two systems.
Similar to single SEB's, increasing tunneling conductance in the junction $b$ 
increases the capacitance $C_b$ resulting in reduced curvature of
the energy parabolas and smaller $\tilde{\Delta}$'s. 
This can be seen as a broadening and shifting of the conductance peaks.
The resulting peak shift is shown in Fig.~\ref{fg:peakshift} as
a function of $\alpha_0^{\rm L,R}$.
Increasing tunneling conductance also results in 
a gate-voltage and temperature dependent renormalization of
$\tilde{\alpha}_{\rm L,R}$ and $\tilde{\alpha}_{\rm M}$ reducing these
quantities.
Figure \ref{fg:fullG_aL} shows the full conductance curves for two choices
of $\alpha_0^{\rm L,R}$. The curious shape of the peaks is a consequence of 
the interplay of two effects: 
at a finite temperature the peak is shifted away from $n_x^*$  
due to unequal $\alpha_0^b$'s, see above, while $\tilde{\alpha}_b$  
reduce the conductance around $n_x^*$, i.e., not quite at the actual maximum.

Before going to the third case of equal $\alpha_0^b$'s, let us 
write down the main results obtained by expanding the RG equations
to $O(\alpha_0^2)$ and make comparison to results in the literature.
The renormalized degeneracy point, $n_x^*$, follows from the condition 
$\tilde{\Delta}(n_x^*)=\Delta^{(0)}(n_x^*)+\Delta^{(1)}(n_x^*)=0$.
In the zero-temperature limit, the RG calculation yields the temperature- 
and $n_x$-independent correction
\begin{eqnarray}
  & \Delta^{(1)} & = \nonumber \\
        & \alpha_0^{\rm L} & \left(2E_{\rm CL}\ln\left|{2E_{\rm CL} \over 
                                        {2E_{\rm CL}-E_{\rm CM}}}\right|
    + E_{\rm CM}\ln\left|{2E_{\rm CL}-E_{\rm CM}\over 
                                        E_{\rm CM}}\right|\right) 
  \nonumber \\ \nonumber \\
      & - & \alpha_0^{\rm M}(4E_{\rm CL}-2E_{\rm CM})\ln2.
\label{eq:Delta(1)}
\end{eqnarray}
For the special
case $E_{\rm CL,R}=E_{\rm CM}$, the prefactors of $\alpha_0^{\rm M}$ 
and $\alpha_0^{\rm L}$ are equal.
This linear $\alpha_0^{\rm M}$-dependence is a new result whereas 
the $\alpha_0^{\rm M}$-dependence equals the results in
Refs.~\onlinecite{Gol-Hal,Gol-Hal_largeN,Mat-Gla-Bar_shift} 
(shown to be independent of $N$ in Ref.~\onlinecite{Gol-Hal_largeN}).
Further examination of the $\alpha_0^{\rm M}$-dependent curve in 
Fig.~\ref{fg:peakshift} in terms of polynomial fitting yields 
the coefficient of the quadratic term within 1-2\% of the analytical result 
Ref.~\onlinecite{Gol-Hal_largeN}. It should be noted that this accuracy 
already requires six correct decimals in calculating $n_x^*$.
The $\alpha_0^{\rm L}$-dependent curve in Fig.~\ref{fg:peakshift}
remains very close to linear up to the largest $\alpha_0^{\rm L,R}$ used here. 
However, if we repeat the calculation with a smaller value of $E_{\rm CM}$ 
the curve only approaches the line $n_x=0.5$ asymptotically.

An expression for the temperature dependence of the peak conductance
may be found by inserting the renormalized $\alpha_0^b$'s to 
Eq.~(\ref{eq:Gseq}). This corresponds to the requirement that at resonance 
$\tilde{\Delta}(n_x^*)=0$ and yields 
\begin{eqnarray}
  G & = & G^{\rm seq} + G^{(2)} \\
    & = & G^{\rm seq}\left[1
    +{B_1(\alpha_0^{\rm L})^2+B_2\alpha_0^{\rm L}\alpha_0^{\rm M}+B_3(\alpha_0^{\rm M})^2
      \over 2\alpha_0^{\rm M} + \alpha_0^{\rm L}}\right] \nonumber
\end{eqnarray}
with the temperature dependent  parameters $B_i$: $B_1 = 2\ln|T/E_{\rm CM}|$, 
$B_2 = 8\ln|T/E_{\rm CM}|$, and $B_3 = \ln|T/E_{\rm CM}|$.
For $\alpha=\alpha_0^{\rm L,R}=\alpha_0^{\rm M}$ we obtain the limiting form
\begin{eqnarray}
  G^{(2)} & \propto & {4\over9}\alpha^2\ln|T/E_{\rm CM}|.
\label{eq:logT}
\end{eqnarray}
In this order, this is the sole source of temperature dependence to 
the maximum value of $G$.
In the more general case $\alpha_0^{\rm L,R}\neq\alpha_0^{\rm M}$,
the peak is not at the point $n_x^*$ and one should use the full form 
for $G$.

\begin{figure}[h]
\epsfxsize=8.0cm
\centerline{\epsffile{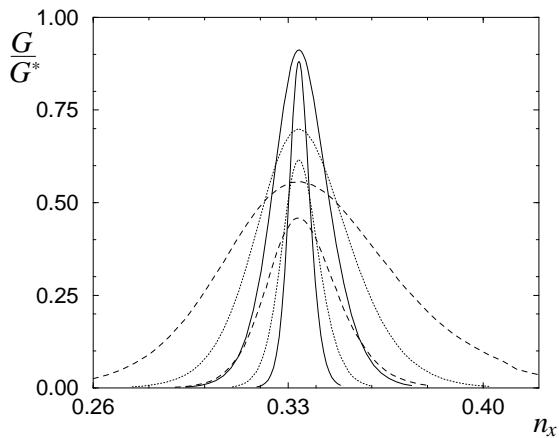}}
\caption{
  The full conductance curve for 
  $\alpha_0\equiv$ $\alpha_0^{\rm L,R}=$$\alpha_0^{\rm M}=0.01,0.05,0.10$ 
  (pairs of solid, dotted and dashed curves, respectively) calculated
  at $\ln(T/E_{\rm C})=-4,-5$ (upper and lower curve of each pair).
  The capacitances are chosen such that the peak does 
  not shift either with temperature or with $\alpha_0$.
  }
\label{fg:fullG_full}
\end{figure}

\begin{figure}[h]
\epsfxsize=8.0cm
\centerline{\epsffile{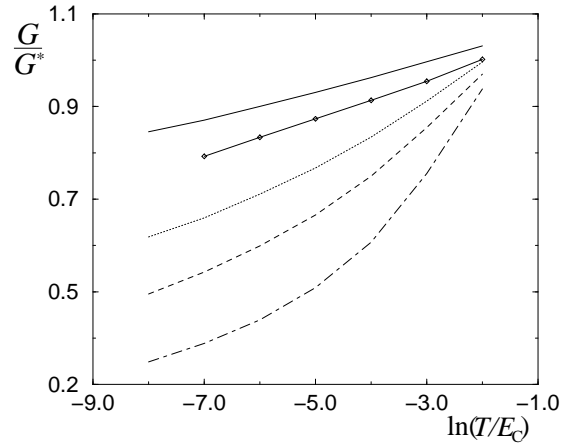}}
\caption{
  Temperature dependence of the maximum conductance for
  $\alpha_0\equiv\alpha_0^{\rm L,R}=\alpha_0^{\rm M}=0.01,0.01^*,0.03,0.05,0.10$ 
  from up to down (* the curve with the diamond symbols is the second-order
  result for $\alpha_0=0.01$). 
  Below $\ln(T/E_{\rm C})=-3$, the curves are of the functional form
  $1/(A+B\ln(T/E_{\rm C}))$.
  The 
  }
\label{fg:GlnT}
\end{figure}

In the complete case, when charge fluctuations are allowed through all
three junctions, the above-discussed effects intermingle in a complicated way. 
Now all the energies $E_{{\rm C}b}$ and tunneling conductances $\alpha_0^b$
are renormalized with some contributions to these depending on gate voltages 
and/or temperature while some others remain constant.  
In the following we consider for convenience the case of all $\alpha_0^b$ 
being equal. 
Even in this case, the peak may be shifted, cf. Eq.~(\ref{eq:Delta(1)}), 
but, in order to see the changes in the peak shapes more easily,
the capacitances are chosen such that $n_x^*$ is independent
of the tunneling conductances.
The choice of capacitances adopted above is close to the special case 
$E_{\rm CL,R}=E_{\rm CM}$ with no shift in the second order and it appears
that in this case there is no shift in higher orders either.
Figure \ref{fg:fullG_full} shows the full conductance curves for
three values of $\alpha_0\equiv\alpha_0^b$ at two temperatures 
in order to distinguish 
between the effects arising from thermal and quantum fluctuations.
The broadening of the peaks with incresing $\alpha_0^b$ is due to 
the overall reduction of $\Delta$'s as well as the gate-voltage dependent 
reduction of $\tilde{\alpha}_b$.
Figure \ref{fg:GlnT} shows the temperature dependence of the maximum conductance
for a few values of $\alpha_0$. 
At low temperatures, the curves resemble the functional form 
$1/(A+B\ln(T/E_{\rm C}))$, with the coefficient $B$ being proportional to
$\alpha_0$. 
This is consistent with the assumption that the full RG calculation 
resums the leading logarithmic terms $\alpha_0\ln(T/E_{\rm C})$.
Figure~\ref{fg:GlnT} also shows the second-order result for $\alpha_0=0.01$
with the slope given by Eq.~(\ref{eq:logT}).
The $\Delta^{(1)}/T$-behaviour of Fig.~\ref{fg:fullG_aM} is absent because
the peak is not shifted and $\Delta^{(1)}(n_x^*)=0$.

\vspace{0.3cm}
\section{Conclusions}

In conclusion, we have evaluated the transport properties
of a metal double-island structure using two approaches:
the real-time diagrammatic technique developed in Ref.~\onlinecite{SS,KSS1}
and the RG-procedure introduced in Ref.~\onlinecite{KS-RG}.
We made no {\it a priori} assumptions concerning the relative magnitudes 
of the tunneling conductances but we assumed
a large number of channels in each junction.
The perturbative (diagrammatic) expansion has been 
performed up to $O(\alpha_0^2)$, 
limiting the results to small values of $\alpha_0$. In this case we obtained 
both the linear and nonlinear response 
of the system for all values of gate voltage. 
In the RG calculation, we considered equilibrium properties of the system
covering temperatures down to zero and
also much larger tunnel conductances than in the perturbation theory.

Due to periodicity of the charging energies, it is sufficient to concentrate 
on the single conductance peak appearing in linear conductance
in the gate-voltage range $0\le n_x\le 0.5$. The height and position
of this peak are found to be renormalized and its shape becomes asymmetric.
In general, we do not need to restrict the relative magnitudes of 
the tunneling conductances $\alpha_0^b$, but two limiting cases can be 
compared to existing work on similar systems.
In the limit of $\alpha_0^M\gg\alpha_0^L=\alpha_0^R$, we 
reproduce the third-order results for the peak position reported by 
Golden and Halperin\cite{Gol-Hal_largeN}.
In the opposite limit, $\alpha_0^M\ll\alpha_0^L=\alpha_0^R$,
the system resembles the model considered by Andrei~{\it et al.}\cite{AZS}.
These authors studied the mapping of a double-island system with one 
conducting channel in all junctions onto a Kondo model.
It was predicted that the system would undergo a phase transition as 
a function of $E_{\rm CM}$, 
the mutual interaction energy for electrons in different islands:
the two conductance peaks around $n_x=e/2$ merge into one 
as $E_{\rm CM}$ is reduced below some finite critical value $E_{\rm CM}^{cr}$.
By putting $\alpha_0^{\rm M}$ small in our calculations for many-channel 
junctions, we interestingly obtain a similar trend of peaks shifting 
towards each other, see Fig.~\ref{fg:peakshift}.

We suggest the following experimental schemes similar to those 
used in Ref.\ \onlinecite{Joyezetal}: 
one could measure the heights both the linear and nonlinear conductance, 
as well as the asymmetric shape of the double-peak structure in 
the nonlinear conductace $G(n_x,V\not=0)$.
The tunneling conductance of a single junction is typically
fixed in metallic structures. 
However, if one could attach an additional gate partially on top of 
a junction itself, one might be able to control the number of conducting 
channels instead of the conductance of separate channels.
This would enable the study of the peak position as a function of 
the tunneling conductances $\alpha_0^b$.

\vspace{0.5cm}
\section{Acknowledgements}

This work has been supported by the ``Deutsche 
Forschungsgemeinschaft'' as part of ``Sonderforschungbereich 195.''
One of us (GS) also gratefully acknowledges the support 
through an A.~v.~Humboldt Research Award of the Academy of Finland
and one (TP) the support through a post-graduate Research 
Fellowship from the Finnish Academy of Science and Letters
as well as support through EU TMR Network ``Dynamics of Nanostructures''.

\appendix

\section{Diagrammatic calculation to $O(\alpha_0^2)$}
\label{app:rates}

In this appendix we present a perturbative calculation of the conductance 
through the double-island system up to the order $O(\alpha_0^2)$.
We also summarize the rules for evaluating the second-order
rates $\Sigma_{\chi,\chi^\prime}^{(2)}$ in the energy representation, i.e.,
in Fourier space.

We begin by inserting both the probabilities and the rates in their
series form, Eq.~(\ref{eq:series}), into the master equation 
Eq.~(\ref{eq:master}) and to Eq.~(\ref{eq:current}) for the current. 
We then pick up the terms of the orders $O(\alpha_0)$ and $O(\alpha_0^2)$.
The lowest-order contributions are of the form $p^{(0)}\Sigma^{(1)}$, 
corresponding to sequential-tunneling processes, while
in the second order two kinds of corrections appear.
Terms such as $p^{(0)}\Sigma^{(2)}$ correspond to two-electron 
tunneling processes including the ``inelastic cotunneling''
processes~\cite{Ave-Naz} but also processes that give rise to 
renormalization of system parameters.  
Terms of the form $p^{(1)}\Sigma^{(1)}$ arise due to changes 
in the occupation probabilities in higher orders.
While away from resonances the ``inelastic cotunneling'' yields the dominant
contribution to the current, 
it was recently shown that close to the resonances
it is crucial to consider all these second-order terms \cite{KSS-cot}.

The master equation, Eq.~(\ref{eq:master}), must hold in each order
and once we know the rates we may iteratively solve for $p^{(k)}$.
Hence, the essential step of the perturbative calculation consists of 
the evaluation of the transition rates $\Sigma^{(1)}$ and $\Sigma^{(2)}$. 
The former are just the Golden rule rates and also the latter
may be evaluated analytically as was done in
Ref.\ \onlinecite{KSS-cot}.
Before going into detail in evaluating the second-order rates, 
we would like to comment on two technical points.
First, at intermediate stages we find it convenient to introduce 
a cutoff procedure to ascertain convergence of integrals.
Any cutoff dependence cancels from all physical quantities, but
only when we consider a set of 12 charge states -- for the values
of gate voltage used here these are
(0,0), (1,0), (0,1), (-1,0), (0,-1), (1,1), 
(2,0), (0,2), (-1,1), (1,-1), (-1,2), and (2,-1)\cite{chargestates}.
If we would not account for the additional charge states,
the resulting quantities would contain terms with linear and 
logarithmic cutoff dependences. 
Second, for convenience, we solve the resulting set of 12 master 
equations numerically as this requires inverting a 12-by-12 matrix.

\subsection{Rates}

The tunneling rates may be visualized in terms of diagrams such as shown
in Fig.~\ref{fg:diagram}, each diagram consisting of the following parts.
The contour (solid line) corresponds to a forward and backward propagator,
the upper and lower lines, respectively. Each part of the propagators
has a charge state $(n_{\rm l},n_{\rm r})$ assigned to it
and it carries the corresponding energy $E_{n_{\rm l},n_{\rm r}}$.
The directed dashed lines correspond to tunneling of single electrons through 
the junctions. These tunneling lines carry an index $b$, corresponding
to the junction, as well as an energy $\omega$ of the tunneling electron.  
The diagrams, that cannot be divided into parts by a vertical cut
through them without cutting one of the tunneling lines, are denoted
irreducible diagrams and they correspond to rates of individual tunneling 
processes.
The tunneling rates between states $\chi$ and $\chi^\prime$
($\chi=n_{{\rm l},\chi},n_{{\rm r},\chi}$) are obtained by summing up 
the rates of all the tunneling processes starting from the state $\chi$ and 
ending up in the state $\chi^\prime$.

An arbitrary diagram may be evaluated as follows.
1. Each tunneling line corresponds to the expression
\begin{equation}
        \alpha_\b^\pm(\omega) = \pm\alpha_0^\b
        {{\omega-\Delta\mu_\b} 
        \over {\exp\left[\pm\beta(\omega-\Delta\mu_\b)\right] - 1}},
\end{equation}
with the plus (minus) sign corresponding to forward (backward)
direction of the line with respect to the contour.
The coefficient $\alpha_0^\b$ is the dimensionless conductance of 
the junction $b$, $\Delta\mu_\b$ is the difference in chemical potentials
across the junction and $\omega$ is the energy of the tunneling electron. 
2. In order to account for all possible tunneling processes, one 
integrates over the energies $\omega$ and $\omega^\prime$ of the two 
tunneling lines.
3. The vertices, i.e., the end points of the tunneling lines, 
lying on the backward (lower) propagator acquire an extra factor $-1$ each.
4. The diagrams can be divided into three parts $i$ such that within each 
of them the energies of the propagators and tunneling lines remain constant.
Each part $i$ yields an energy denominator $1/(\Delta E_i+i0^+)$ 
corresponding to the energy of the intermediate virtual state: 
$\Delta E_i$ equals the difference of all energies of the lines 
(propagators and tunnelng lines) directed 
to the left and those directed to the right. 
%
%
The rate of a given process is the imaginary part of the expression
obtained by evaluating the corresponding diagram. 
For example, we obtain for the third diagram in Fig.~\ref{fg:diagram}
\begin{eqnarray}
     -i\cdot {\rm Im}& \cdot & \int d\omega^\prime
                \alpha_{\rm M}^- (\omega^\prime)
                \int d\omega \alpha_{\rm R}^+(\omega) 
                \cdot {1\over \omega^\prime+\Delta E_1+i0^+} \nonumber \\
      & \cdot & {1\over \omega+\omega^\prime+\Delta E_2+i0^+}
        \cdot {1\over \omega+\Delta E_3+i0^+}. 
\label{eq:integral}
\end{eqnarray}
These integrals may be evaluated analytically along the lines of 
Ref.\ \onlinecite{KSS-cot}.
It should be noted, that the use of these rules implicitly assumes a large 
number $N$ of transverse channels in each tunnel junction.

\subsection{Results}

\begin{figure}[h]
\epsfxsize=7.5cm
\centerline{\epsffile{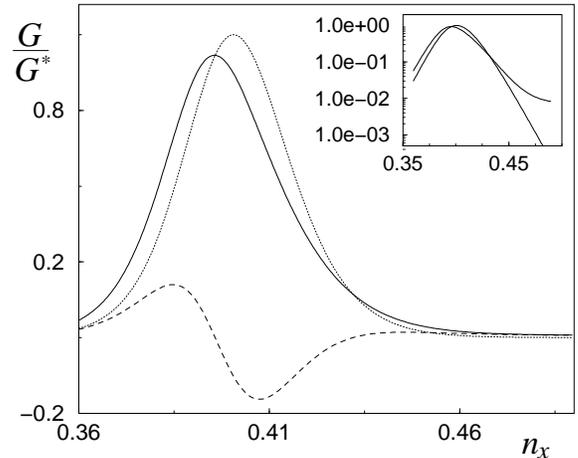}}
\caption{
  Sequential tunneling (dotted), cotunneling (dashed), and the total
  conductance (solid) as the sum of the two as obtained from the diagrammatic
  calculation. The logarithmic scale in the inset emphasizes the functional 
  difference of tails of the sequential-tunneling and total second-order 
  conductances. The parameters used in the plots are
  $\alpha_0^{\rm L,R}=\alpha_0^{\rm M}=0.01$ 
  and $T=0.03E_{\rm C}$.
  The junction capacitances are now chosen unequal 
  $C_{\rm M}={1\over3}C_{\rm L,R}$ but still two orders of magnitude larger 
  than the gate capacitances. 
  This results in $E_{\rm CM}={1\over2}E_{\rm CL,R}={4\over5}E_{\rm C}$.
  }
\label{fg:noshift}
\end{figure}

Conductance away from the peak is mainly given by the inelastic cotunneling 
processes and cannot be described by equilibrium calculations such
as the RG analysis in the main text.
For certain values of the gate voltages, the lowest energy state is
doubly degenerate, see Fig.~\ref{fg:honeycomb}. 
In these regions we obtain a finite second-order contribution as, e.g.,
on the right-hand side of the peak in Fig.~\ref{fg:noshift}.
The simplest processes that carry current through 
the whole double-island system are of the order $O(\alpha_0^3)$
and hence beyond the present considerations.
For this reason we only obtain the exponential decay 
on the left side of the peak in Fig.~\ref{fg:noshift}.
Comparison between the present result and the RG calculation displays 
good agreement for small tunneling conductances thus supporting 
the assumption made (in RG) for the functional form of the conductance curve.

\begin{figure}
\epsfxsize=7.5cm
\centerline{\epsffile{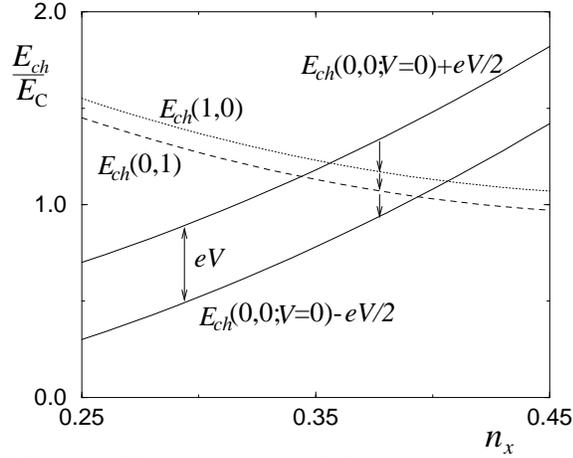}}
\caption{
  Charging energy of the states (0,0), (1,0) and (0,1) when a finite symmetric 
  bias voltage $V$ is applied across the system, i.e., $V_{\rm L}=-V_{\rm R}$. 
  The relevant current-carrying cycle of sequential-tunneling processes  
  is depicted with arrows.
  }
\label{fg:parabolaswithV}
\end{figure}

\begin{figure}[h]
\epsfxsize=8.5cm
\centerline{\epsffile{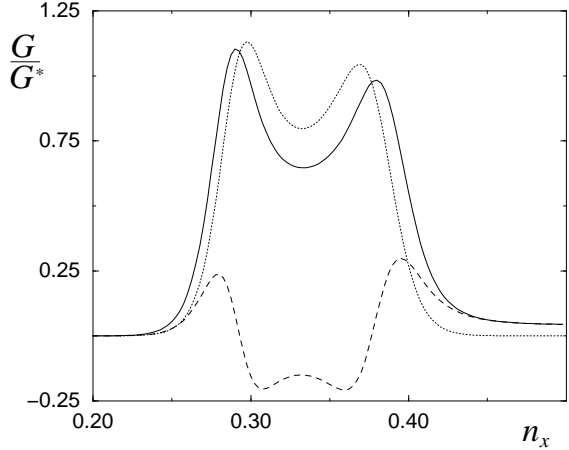}}
\caption{
  Nonlinear conductance through the double-island structure:
  sequential tunneling (dotted) and cotunneling (dashed) contributions,
  and the total conductance (solid).
  The peak structure is asymmetric due to the asymmetry in charging energies
  with respect to the peak positions. 
  The parameters in the figure are $eV=0.4E_{\rm CM}$, $T=0.03E_{\rm CM}$,
  $E_{\rm CM}=E_{\rm CL,R}={4\over3}E_{\rm C}$ and all $\alpha_0^b$=0.01. 
  }
\label{fg:nonlinearG} 
\end{figure}

A finite transport voltage applied across the system changes
the charging energies as shown in Fig.~\ref{fg:parabolaswithV}. 
At low temperature and when we only include sequential tunneling,
the differential conductance $dI/dV$ displays a double peak somewhat 
similar to that found in a SET but now the peaks are slightly asymmetric,
see Fig.~\ref{fg:nonlinearG}. 
For higher values of $\alpha_0$'s the double-peak structure may be shifted 
as a whole, similar to the shift of the single peak in linear conductance. 
In order to inspect any additional effects resulting from the finite voltage, 
we have chosen the parameters in Fig.~\ref{fg:nonlinearG} such that 
the center of the two peaks is not shifted. 
In this case, the peaks are shifted away from each other and 
the asymmetry of their heights and widths is enhanced.
The different shapes of the tails are of the 
same origin as discussed above for the linear conductance, i.e., 
the asymmetric structure of the charging energies.

\section{RG formalism}
\label{app:RG}

In this Appendix we present central formulas of the RG approach
applied in the second part of the text. The details may be found in
\cite{KS-RG}.

We study the renormalization of charging energies $E_{n_{\rm L},n_{\rm R}}$ 
and tunnel conductances $\alpha_{n_{\rm L},n_{\rm R}}^b$. 
As the subscripts already indicate, the various $E$'s and $\alpha$'s 
renormalize in different ways for different charge states.  
For the following, let us define the differences of charging energies
more generally,
\begin{eqnarray}
  \Delta_{n_{\rm l},n_{\rm r}}^{{\rm L}\sigma} & = & E_{ch}(n_{\rm l}+\sigma,n_{\rm r})
  -E_{ch}(n_{\rm l},n_{\rm r}), \nonumber \\
  \Delta_{n_{\rm l},n_{\rm r}}^{{\rm R}\sigma} & = & E_{ch}(n_{\rm l},n_{\rm r}+\sigma)
  -E_{ch}(n_{\rm l},n_{\rm r}),\;\;{\rm and} 
\label{eq:Deltas} \\
  \Delta_{n_{\rm l},n_{\rm r}}^{{\rm M}\sigma} & = & E_{ch}(n_{\rm l},n_{\rm r}+\sigma)
  -E_{ch}(n_{\rm l}+\sigma,n_{\rm r}) \nonumber
\end{eqnarray}
where $\sigma=\pm1$.
The quantities corresponding to those used in the diagrammatic expansion are
$\alpha_{(0,0)}^{\rm L}=\alpha_0^{\rm L}$,  
$\alpha_{(0,0)}^{\rm R}=\alpha_0^{\rm R}$, 
$\alpha_{(1,0)}^{\rm M}=\alpha_0^{\rm M}$, and
$\Delta_{(0,0)}^{b+}=\Delta_{\rm b}$.

We study the RG flow of the charging energies directly but
the tunneling conductances are obtained from the auxiliary functions $g$
defined as
\begin{eqnarray}
  e^{\pm i\hat{\phi}_l} & \rightarrow &
    \sum_{n_{\rm L},n_{\rm R}} g_{n_{\rm L},n_{\rm R}}^{\rm L\pm}
    |n_{\rm L}\pm1, n_{\rm R}\rangle\langle n_{\rm L},n_{\rm R}| \nonumber \\
  e^{\mp i(\hat{\phi}_l-\hat{\phi}_r)} & \rightarrow &
    \sum_{n_{\rm L},n_{\rm R}} g_{n_{\rm L},n_{\rm R}}^{\rm M\pm}
    |n_{\rm L}\pm1, n_{\rm R}\mp1\rangle\langle n_{\rm L},n_{\rm R}| \\
  e^{\pm i\hat{\phi}_r} & \rightarrow &
    \sum_{n_{\rm L},n_{\rm R}} g_{n_{\rm L},n_{\rm R}}^{\rm R\pm}
    |n_{\rm L}, n_{\rm R}\pm1\rangle\langle n_{\rm L},n_{\rm R}|. \nonumber
\end{eqnarray}
These are introduced to parametrize the renormalization of $\alpha_0$'s
\begin{eqnarray}
  \alpha_{n_{\rm L},n_{\rm R}}^{\rm L} & = &
           \alpha_0^{\rm L}\; g_{n_{\rm L},n_{\rm R}}^{\rm L+}
                           \; g_{n_{\rm L}+1,n_{\rm R}}^{\rm L-}\nonumber\\
  \alpha_{n_{\rm L},n_{\rm R}}^{\rm R} & = &
           \alpha_0^{\rm R}\; g_{n_{\rm L},n_{\rm R}}^{\rm R+}
                           \; g_{n_{\rm L},n_{\rm R}+1}^{\rm R-} \\
  \alpha_{n_{\rm L},n_{\rm R}}^{\rm M} & = &
           \alpha_0^{\rm M}\; g_{n_{\rm L},n_{\rm R}}^{\rm M+}
                           \; g_{n_{\rm L}-1,n_{\rm R}+1}^{\rm M-}.\nonumber
\end{eqnarray}
Initially all $g$'s equal unity corresponding to the bare $\alpha$'s.

As an example of the full RG equations in
the equilibrium case we have for the energies 
\begin{eqnarray}
  & {d\over dl} & \bar{\Delta}_{n_{\rm L},n_{\rm R}}^{\rm L} = 
  {d\over dl}(\bar{E}_{n_{\rm L}+1,n_{\rm R}}-\bar{E}_{n_{\rm L},n_{\rm R}})
  = \bar{\Delta}_{n_{\rm L},n_{\rm R}}^{\rm L}   \nonumber\\
  & + & i\sum_{\chi,\chi^\prime} (-1)^\chi \left\{
    \alpha_{n_{\rm L}-\chi+\chi^\prime,n_{\rm R}}^{\rm L}
    \exp[(-1)^{\chi^\prime} i \bar{\Delta}_{n_{\rm L}-\chi+\chi^\prime,n_{\rm R}}^{\rm L}] \right. \nonumber \\
  & - & \alpha_{n_{\rm L}+\chi,n_{\rm R}-\chi^\prime}^{\rm R}
    \exp[-(-1)^{\chi^\prime} i \bar{\Delta}_{n_{\rm L}+\chi,n_{\rm R}-\chi^\prime}^{\rm R}] \\
  & - & \left. \alpha_{n_{\rm L}+1-\chi+\chi^\prime,n_{\rm R}-\chi}^{\rm M}
    \exp[-(-1)^{\chi^\prime} i \bar{\Delta}_{n_{\rm L}+1-\chi+\chi^\prime,n_{\rm R}-\chi}^{\rm M}]\right\} 
\label{eq:RG_delta-general}
\end{eqnarray}
The summation indices $\chi$ and $\chi^\prime$ take values zero and one.
All quantities with a bar on top of them contain the time $t_c$ corresponding 
to the high-energy cutoff of the theory as, e.g., 
$\bar{E}_{n_{\rm L},n_{\rm R}}=E_{n_{\rm L},n_{\rm R}}t_c$.
The parameter $l$ is defined as $l=\ln(t_c/t_c^0)$ where
$t_c^0$ is the initial cutoff time.
At the end, we take the limit $t_c^0 \rightarrow 0$ and
integrate the equations to $t_c \rightarrow {\rm min}\{1/\Delta,1/T\}$.

The equations for charging energies are coupled with the flow equations
for the $g_{n_{\rm L},n_{\rm R}}^{b\sigma}$'s ($\sigma=\pm1$)
which follow equations such as
\begin{eqnarray}
  & {d\over dl} & g_{n_{\rm L},n_{\rm R}}^{{\rm L}\sigma} =
  \sum_{\sigma^\prime=\pm1} \nonumber \\ 
  &\cdot& \left\{ \alpha_0^{\rm L} 
    g_{n_{\rm L}+\sigma-\sigma^\prime,n_{\rm R}}^{{\rm L}\sigma^\prime}
    g_{n_{\rm L}-\sigma^\prime,n_{\rm R}}^{{\rm L}\sigma}
    g_{n_{\rm L},n_{\rm R}}^{{\rm L},-\sigma^\prime}\right.\nonumber\\
  &&\;\;\;\;\;\;\;\;\;\;\;\;\;\;\;\;\;  {{\exp
 (i\bar{\Delta}_{n_{\rm L}+\sigma-\sigma^\prime,n_{\rm R}}^{{\rm L}\sigma^\prime})
 - \exp
 (i\bar{\Delta}_{n_{\rm L}-\sigma^\prime,n_{\rm R}}^{{\rm L}\sigma^\prime})}
 \over {
   i(\bar{\Delta}_{n_{\rm L}+\sigma-\sigma^\prime,n_{\rm R}}^{{\rm L}\sigma^\prime}
   - \bar{\Delta}_{n_{\rm L}-\sigma^\prime,n_{\rm R}}^{{\rm L}\sigma^\prime})}}
   \nonumber \\
  & - & \alpha_0^{\rm L} 
    g_{n_{\rm L},n_{\rm R}}^{{\rm L}\sigma}
    g_{n_{\rm L}-\sigma^\prime,n_{\rm R}}^{{\rm L}\sigma^\prime}
    g_{n_{\rm L},n_{\rm R}}^{{\rm L},-\sigma^\prime}
    \exp(i\bar{\Delta}_{n_{\rm L}\sigma^\prime,n_{\rm R}}^{{\rm L}\sigma^\prime})
    \\
    & + & \alpha_0^{\rm R} 
    g_{n_{\rm L}+\sigma,n_{\rm R}-\sigma^\prime}^{{\rm R}\sigma^\prime}
    g_{n_{\rm L},n_{\rm R}-\sigma^\prime}^{{\rm L}\sigma}
    g_{n_{\rm L},n_{\rm R}}^{{\rm R},-\sigma^\prime}\nonumber\\
  &&\;\;\;\;\;\;\;\;\;\;\;\;\;\;\;\;\;
    {{\exp
 (i\bar{\Delta}_{n_{\rm L}+\sigma,n_{\rm R}-\sigma^\prime}^{{\rm R}\sigma^\prime})
 - \exp
 (i\bar{\Delta}_{n_{\rm L},n_{\rm R}-\sigma^\prime}^{{\rm R}\sigma^\prime})}
 \over {
i(\bar{\Delta}_{n_{\rm L}+\sigma,n_{\rm R}-\sigma^\prime}^{{\rm R}\sigma^\prime}
- \bar{\Delta}_{n_{\rm L},n_{\rm R}-\sigma^\prime}^{{\rm R}\sigma^\prime})}}
  \nonumber \\
  & - & \alpha_0^{\rm R} 
    g_{n_{\rm L},n_{\rm R}}^{{\rm L}\sigma}
    g_{n_{\rm L},n_{\rm R}-\sigma^\prime}^{{\rm R}\sigma^\prime}
    g_{n_{\rm L},n_{\rm R}}^{{\rm R},-\sigma^\prime}
    \exp(i\bar{\Delta}_{n_{\rm L},n_{\rm R}-\sigma^\prime}^{{\rm R}\sigma^\prime})
    \nonumber \\
    & + & \alpha_0^{\rm M} 
    g_{n_{\rm L}+\sigma+\sigma^\prime,n_{\rm R}-\sigma^\prime}^{{\rm M}\sigma^\prime}
    g_{n_{\rm L}+\sigma^\prime,n_{\rm R}-\sigma^\prime}^{{\rm L}\sigma}
    g_{n_{\rm L},n_{\rm R}}^{{\rm M},-\sigma^\prime}\nonumber\\
  &&\;\;\;\;\;\;\;\;\;\;\;\;\;\;\;\;\;
    {{\exp
 (i\bar{\Delta}_{n_{\rm L}+\sigma+\sigma^\prime,n_{\rm R}-\sigma^\prime}^{{\rm M}\sigma^\prime})
 - \exp
 (i\bar{\Delta}_{n_{\rm L}+\sigma^\prime,n_{\rm R}-\sigma^\prime}^{{\rm M}\sigma^\prime})}
 \over {
i(\bar{\Delta}_{n_{\rm L}+\sigma+\sigma^\prime,n_{\rm R}-\sigma^\prime}^{{\rm M}\sigma^\prime}
- \bar{\Delta}_{n_{\rm L}+\sigma^\prime,n_{\rm R}-\sigma^\prime}^{{\rm M}\sigma^\prime})}}
  \nonumber \\
  & - & \left.\alpha_0^{\rm M}  
    g_{n_{\rm L},n_{\rm R}}^{{\rm L}\sigma}
    g_{n_{\rm L}+\sigma^\prime,n_{\rm R}-\sigma^\prime}^{{\rm M}\sigma^\prime}
    g_{n_{\rm L},n_{\rm R}}^{{\rm M},-\sigma^\prime}
    \exp(i\bar{\Delta}_{n_{\rm L}+\sigma^\prime,n_{\rm R}-\sigma^\prime}^{{\rm M}\sigma^\prime})
  \right\}.\nonumber
\end{eqnarray}    
The full solution to these can be obtained via numerical integration.

Expanding the above equations up to $O(\alpha_0^2)$ yields the leading
corrections to $\Delta$'s and $\alpha$'s. 
The resulting equations are decoupled such that the equations for
second-order quantities only depend on the respective bare quantities.
For this reason the calculations can be carried out analytically.
For the $\Delta$'s we arrive at 
\begin{eqnarray}
  {d\over dt_c}\Delta_{n_{\rm L},n_{\rm R}}^{b^\prime(1)}
  & = & \\
  & - & {i\over t_c^2} \sum_{b^\prime\sigma} \alpha_0^{b^\prime}
  e^{-i\Delta_{n_{\rm L},n_{\rm R}}^{b^\prime\sigma(0)}t_c}
  - (n_b \rightarrow n_{b+1}). \nonumber
\end{eqnarray}
The notation means that the second part of the equation 
equals the first part up to the indicated replacements of 
$n_b$'s. 
Here $b,b^\prime\in\{{\rm L,\;R}\}$ and 
$\Delta_{n_{\rm L},n_{\rm R}}^{{\rm M}\sigma^\prime}
=\Delta_{n_{\rm L}-\sigma^\prime,n_{\rm R}}^{{\rm R}\sigma^\prime}
-\Delta_{n_{\rm L}-\sigma^\prime,n_{\rm R}}^{{\rm L}\sigma^\prime}$.
These equations can be solved to yield
\begin{eqnarray}
&& \Delta_{n_{\rm L},n_{\rm R}}^{b(1)} (t_c) = 
   + i\sum_{b^\prime\sigma} \alpha_0^{b^\prime} \left\{ 
    {\exp(-i\Delta_{n_{\rm L},n_{\rm R}}^{{\rm L}\sigma(0)}t_c) \over t_c}
    \right.\nonumber\\
&& + i\Delta_{n_{\rm L},n_{\rm R}}^{{\rm L}\sigma(0)}
  {\rm Ci}\left(\left|\Delta_{n_{\rm L},n_{\rm R}}^{{\rm L}\sigma(0)}\right|t_c\right)
    + \left|\Delta_{n_{\rm L},n_{\rm R}}^{{\rm L}\sigma(0)}\right|
    {\rm Si}\left(\left|\Delta_{n_{\rm L},n_{\rm R}}^{{\rm L}\sigma(0)}\right|t_c\right)
    \nonumber\\
&& - (n_b\rightarrow n_b+1) \bigg\} - (t_c \rightarrow t_c^0)
\end{eqnarray}
The functions Ci($x$) and Si($x$) are cosine and sine integrals.

The functions $g^{(1)}(t_c)$ take the form as, e.g., 
\begin{eqnarray}
  & g_{n_{\rm L},n_{\rm R}}^{b\sigma(1)} & (t_c) = \sum_{r^\prime\sigma^\prime}
  \alpha_0^{b^\prime}\nonumber\\
  & \cdot & \left\{
    \exp(-i\Delta_{n_{\rm L},n_{\rm R}}^{b^\prime\sigma^\prime(0)}t_c)
    {{\exp(-i\sigma\sigma^\prime c_{rr^\prime}t_c) - 1}
      \over i\sigma\sigma^\prime c_{rr^\prime}t_c} \right. \nonumber \\
  & + & \left(1+\sigma\sigma^\prime
    {\Delta_{n_{\rm L},n_{\rm R}}^{b^\prime\sigma^\prime(0)}\over c_{rr^\prime}}
    \right) \\
  & \cdot & \left[
    {\rm Ci}\left(\left|\Delta_{n_{\rm L},n_{\rm R}}^{b^\prime\sigma^\prime(0)}
        + \sigma\sigma^\prime c_{bb^\prime}\right|t_c\right)
    - {\rm Ci}\left(\left|\Delta_{n_{\rm L},n_{\rm R}}^{b^\prime\sigma^\prime(0)}
       \right|t_c\right)\right.\nonumber \\
  & - & i{\rm Si}\left(\left(\Delta_{n_{\rm L},n_{\rm R}}^{b^\prime\sigma^\prime(0)}
        + \sigma\sigma^\prime c_{bb^\prime}\right)t_c\right)
    + i{\rm Si}\left(\left(\Delta_{n_{\rm L},n_{\rm R}}^{b^\prime\sigma^\prime(0)}
       \right)t_c\right) \nonumber \\
   & - & (t_c\rightarrow t_c^0) \big]\Big\}.\nonumber
\end{eqnarray}
If $\Delta=0$ the replacement
\begin{equation}
{\rm Ci}\left(\left|\Delta_{n_{\rm L},n_{\rm R}}^{b^\prime\sigma^\prime(0)}
       \right|t_c^0\right) \rightarrow \gamma 
     + \ln\left|\Delta_{n_{\rm L},n_{\rm R}}^{b^\prime\sigma^\prime(0)}\right|
   + \ln(t_c) 
\end{equation}
has to be made.
The summations over $\sigma$ and $\sigma^\prime$ run over $\pm1$ and
the parameters $c_{b,b^\prime}$ are given by
\begin{eqnarray}
  c_{\rm LL} & = & 2E_{\rm CL},\; c_{\rm RR} = 2E_{\rm CR} \nonumber \\
  c_{\rm MM} & = & 2(E_{\rm CL}+E_{\rm CR}-E_{\rm CM})   \nonumber \\
  c_{\rm ML} & = & c_{\rm LM}=E_{\rm CM}-2E_{\rm CL}   \\ 
  c_{\rm LR} & = & c_{\rm RL}=E_{\rm CM} \nonumber \\ 
  c_{\rm MR} & = & c_{\rm RM}=2E_{\rm CR}-E_{\rm CM}. \nonumber 
\end{eqnarray}
The quantities on the right-hand sides are the charging energies 
$E_{{\rm C}b}$ from Eq.(\ref{eq:Ech}).


\begin{references}
\bibitem{sigmaplus}
The rate $\Sigma_{\chi^\prime,\chi}^{\b+}$ for processes which
change the state from $\chi^\prime$ to $\chi$ has the following diagrammatic 
representation:
the rightmost tunneling line corresponds to the barrier $\b$, 
and it is an outgoing (incoming)
one if the rightmost vertex lies on the upper (lower) propagator.
As an example, the three processes depicted in Fig.\ref{fg:diagram}
contribute to rates $\Sigma_{(0,0),(1,0)}^{\rm L+}$,
$\Sigma_{(1,0),(0,1)}^{\rm R-}$, and $\Sigma_{(0,1),(0,1)}^{\rm L+}$,
respectively.
\bibitem{expansion}
The terms $\Delta^{(1)}(n_x)$ and $\alpha_b^{(2)}$ are the leading corrections
to the bare parameters and are of the order $O(\alpha_0)$
and $O(\alpha_0^2)$, respectively.
(Note: $\alpha_0^b$ is of 1st and its correction of 2nd order, 
while $\Delta^{(0)}$ is of 0th and its correction of 1st order.)
\bibitem{chargestates}
The set of states required for the completeness of the theory 
in second order can be found as follows: take the charge states 
reachable via sequential tunneling and include all states
which may be reached from these via an arbitrary process of
the 1st order -- these states are virtually occupied
during tunneling processes.
\end{references}
\end{document}